\documentclass[preprint]{ptephy_v1}

\preprintnumber{OU-HET-900} 

\usepackage{amsmath} 
\usepackage{amsthm} 
\usepackage{graphics} 
\usepackage{braket}
\usepackage{mathrsfs} 

\numberwithin{equation}{section}
\def\Rnum#1{\uppercase\expandafter{\romannumeral #1}} 

\begin{document}

\title{Six-dimensional regularization of chiral gauge theories}


\author{Hidenori Fukaya}
\author{Tetsuya Onogi}
\author{Shota Yamamoto}
\author{Ryo Yamamura}
\affil{Department of Physics, Osaka University, Toyonaka, Osaka 560-0043 Japan
\email{hfukaya@het.phys.sci.osaka-u.ac.jp, onogi@phys.sci.osaka-u.ac.jp, syamamoto@het.phys.sci.osaka-u.ac.jp, ryamamura@het.phys.sci.osaka-u.ac.jp}
}




\begin{abstract}%
We propose a regularization of four dimensional chiral gauge theories using six-dimensional Dirac fermions.
In our formulation, we consider two different mass terms having 
domain-wall profiles in the fifth and the sixth directions, respectively.
A Weyl fermion appears as a localized mode at the junction of two different domain-walls.
One domain-wall naturally exhibits the Stora-Zumino chain of the anomaly descent equations,
starting from the axial $U(1)$ anomaly in six-dimensions to the gauge anomaly in four-dimensions.
Another domain-wall implies a similar inflow of the global anomalies.
The anomaly free condition is equivalent to requiring that the axial $U(1)$ anomaly 
and the parity anomaly are canceled among the six-dimensional Dirac fermions.
Since our formulation is based on a massive vector-like fermion determinant,
a non-perturbative regularization will be possible on a lattice.
Putting the gauge field at the four-dimensional junction and extending it 
to the bulk using the Yang-Mills gradient flow,
as recently proposed by Grabowska and Kaplan, 
we define the four-dimensional path integral of the target chiral gauge theory.
\end{abstract}

\subjectindex{B00, B01, B05, B30, B31, B38}

\maketitle

\section{Introduction}
\label{sec:Intro}

Defining chiral fermions on a lattice has been a big challenge since 
Nielsen and Ninomiya \cite{Nielsen:1980rz, Nielsen:1981xu} proved
the no-go theorem about chiral symmetry without unphysical doublers.
The problem was partly solved in the formulation of vector-like gauge theories
on a lattice \cite{Kaplan:1992bt,Hasenfratz:1993sp, Neuberger:1997fp, Neuberger:1998wv}.
It has been, however, still difficult to nonperturbatively realize chiral gauge symmetry.
To construct chiral gauge theories,
one must separate the positive and negative chiral modes.
This process usually violates the gauge symmetry, 
and one has to find gauge non-invariant counter-terms\footnote{
The counter-terms are non-perturbatively given for
$U(1)$ \cite{Luscher:1998du} and $SU(2)\times U(1)$ \cite{Kikukawa:2000kd} gauge groups. 
} to recover it even when 
the target theory is anomaly-free \cite{Narayanan:1994gw, Bock:1997vc, Luscher:1999un, Bar:2000qa, Suzuki:2000ii, Kikukawa:2001mw, Kadoh:2007xb}.

Recently, an interesting approach was proposed by Grabowska and Kaplan \cite{Grabowska:2015qpk},
in the 5-dimensional domain wall fermion formulation.
Keeping the 4-dimensional gauge invariance,
they succeeded in coupling the gauge fields to only one chiral mode on the domain wall.
This was made possible by turning off the gauge fields near the anti-domain wall
using the Yang-Mills gradient flow \cite{Luscher:2010iy}.
In their new approach, one can distinguish the 
anomalous and nonanomalous theories by the presence and absence
of the Chern-Simons (CS) term in the bulk of the 5-dimensional space.
If the CS term exists, then 4-dimensional gauge invariance is
not closed on the domain wall alone and the gauge current flows in the extra dimension.
Thus, no consistent 4-dimensional local effective theory exists in the low-energy limit.
The extra dimension plays a more important role than that for vector-like formulations,
as it converts the problem of gauge anomaly into 
the problem of parity invariance (broken by the CS term) in 5-dimensions 
\cite{Callan:1984sa, Naculich:1987ci, Chandrasekharan:1993ag}.

The importance of an extra dimension was also discussed
in studies on global anomalies \cite{Dai:1994kq, Witten:2015aba}. 
It was shown that the global anomaly can be formulated as
the complex phase of the bulk 5-dimensional theory,
which has 4-dimensional target (massless) fermions on its boundary.
They then claimed a more strict definition of the global anomalies:
not only on the mapping torus (on which the $SU(2)$ global anomaly was shown \cite{Witten:1982fp}
), but also {\it any} fermion determinant on a 5-dimensional compact manifold, 
must be real and positive, otherwise the theory becomes anomalous when it has a 4-dimensional boundary.  
The extra dimension is essential since this new notion of anomaly
can never be understood within 4-dimensional space alone.

It would therefore be interesting to consider both the perturbative gauge anomaly and the global anomaly 
at the same time in the new formulation on higher dimensional space-time, 
which was not discussed in Ref.~\cite{Grabowska:2015qpk}.
However, we notice that the extra dimension for the perturbative anomaly in Ref.~\cite{Grabowska:2015qpk} 
and that for the global anomaly in Refs.~\cite{Dai:1994kq, Witten:2015aba} are quite different.
For the former domain wall fermion formulation, the 5th direction is introduced
to separate the left- and right-handed modes.
On the other hand, the extra dimension for the global anomaly
is introduced as a one-parameter family of the fermion determinant phase 
where the chiral fermions are already put on a 4-dimensional space.
It is then natural to speculate that chiral fermions may need two extra directions, 
or in total 6 dimensions, to be formulated.

The relation of 4-dimensional Weyl fermions to 6-dimensional space-time is not a new 
idea but can be found in the literature.
In Refs.~\cite{AlvarezGaume:1985di, DellaPietra:1986qd}, it was shown that
the phase of the Weyl fermion determinant can be given by the 
$\eta$-invariant of a Dirac operator extended in 5 dimensions.
However, this $\eta$-invariant needs a variation with respect to 
another one-parameter family (originally denoted by $u$; 
see also Ref.~\cite{Kaplan:1995pe}
) and we integrate it over a finite range from 0 to 1. 
This fact implies that the phase of the Weyl fermion determinant needs
two parameters to be well defined.

A more direct hint of the 6th dimension was well known 
as the anomaly descent equations by Stora \cite{Stora:1983ct} 
and Zumino \cite{Zumino:1983ew, Zumino:1983rz}. 
They showed that the 4-dimensional (consistent) gauge anomaly 
is obtained uniquely from the 6-dimensional axial $U(1)_A$ anomaly
up to an overall constant. 
Soon after, Alvarez-Gaum\'e and Ginsparg \cite{AlvarezGaume:1983cs} 
and Sumitani \cite{Sumitani:1984ed} proved that 
this over-all constant must be unity, and the 4-dimensional gauge anomaly
originates from the index theorem in 6 dimensions \cite{Atiyah:1984tf}.
There has, however, been no {\it theory} that reproduces these anomaly descent equations
proposed in the literature.

In this work, we formulate a vector-like 6-dimensional
Dirac fermion system in which Weyl fermions are localized at 
the junction of two different kinds of domain walls.
One domain wall is made in a conventional way, giving
a kink mass (let us take this term in the 6th direction) to the fermions.
Another domain wall is made by giving a kink structure in the 5th direction 
to a background operator which is invariant under $U(1)_A$ rotation.
In a sense, our system is a {\it doubly} gapped 6-dimensional topological insulator.
Apart from the domain walls, the Dirac fermions are gapped by
two types of {\it masses} having different quantum numbers.
Each of the domain walls eliminates one mass term from the boundary modes,
and a gapless mode or our target Weyl fermion appears only at the domain wall junction.

As will be shown in this paper, these two domain walls play different roles
in the anomaly cancellations.
The conventional mass domain wall (we call it the $M$ domain wall)
converts the 6-dimensional $U(1)_A$ anomaly into the CS term,
or the 5-dimensional parity anomaly on it. 
The CS term breaks the gauge symmetry at the domain wall junction,
which is absorbed by the gauge anomaly of the Weyl fermion.
Namely, the $M$ domain wall mediates the perturbative anomaly inflow
\begin{eqnarray}
&\mbox{6D $U(1)_A$ anomaly}& \nonumber\\
&\updownarrow& \nonumber\\
&\mbox{5D parity anomaly (CS term)}&\nonumber\\
&\updownarrow&\nonumber\\
&\mbox{4D (perturbative) gauge anomaly}.&
\end{eqnarray}
On the other hand, another domain wall ($\mu$ domain wall) is only sensitive to 
the zero modes which {\it cannot} appear in the index theorem of the $U(1)_A$ symmetry.
In fact, the fermion modes localized on this domain wall 
produces an {\it almost} real determinant
(except those from the domain wall junction) 
and sensitive to the flips of sign,
or mod-two types of the anomaly.
This is true even when the perturbative anomaly is absent.
Therefore, we conjecture that the anomaly mediated by this $\mu$ domain wall
corresponds to a kind of global anomaly.
For the fundamental representation of the $SU(2)$ group, e.g.,
we will show that this anomaly inflow is consistent with a ladder of the mod-two indices:
\begin{eqnarray}
&\pi_5(SU(2))=\mathbb{Z}_2&\nonumber\\
&\updownarrow&\nonumber\\
&\pi_4(SU(2))=\mathbb{Z}_2&\nonumber\\
&\updownarrow&\nonumber\\
&\mbox{4D}\;\;\eta\mbox{-invariant},&
\end{eqnarray}
where the latter part is well known in Ref.~\cite{Witten:1982fp}
but the former homomorphism of $\pi_5(SU(2))$ and $\pi_4(SU(2))$ is not discussed in the literature (on physics).
The two anomaly inflows finally meet at the junction of the domain walls
and determine the perturbative and global anomalies of the Weyl fermion sitting there.

Then the anomaly-free condition is equivalent to requiring
the 6-dimensional theory to be insensitive to both of
the $U(1)_A$ index and the exotic zero modes.
Since these zero modes flip the sign of the fermion determinant, 
the bulk part of the anomaly-free set of fermion determinants 
becomes real and positive (at least in the continuum limit).
The 4-dimensional boundary modes, on the other hand, can have their own complex phases. 

Since our formulation is a Dirac fermion system with vector-like masses in the bulk,
it is natural to assume that nonperturbative lattice regularization is available, 
using a simple Wilson Dirac operator.
Putting the gauge fields on
the junction of the domain walls, and extending it to the 5th and 6th directions
using the Yang-Mills gradient flow,
as proposed in Ref.~\cite{Grabowska:2015qpk}, the 4-dimensional gauge invariance is maintained.

The rest of the paper is organized as follows.
In Sec.~\ref{sec:parity}, we explain how to distinguish 
the Dirac zero modes originating from the $U(1)_A$ anomaly and
those related to the parity anomaly. 
Then we construct the 6-dimensional Dirac fermion system in the continuum theory
and show how the two kinds of anomaly ladders are realized in Sec.~\ref{sec:formulation}.
In Sec.~\ref{sec:anomalyfree}, we discuss the anomaly-free condition.
According to the more strict definition of the global anomaly \cite{Dai:1994kq, Witten:2015aba},
the anomaly-free condition is nontrivial for our target theory 
on a 4-dimensional torus
(which is an essential requirement for lattice regularizations). 
In Sec.~\ref{sec:4dtheory} we discuss how to implement 
the gauge fields localized at the domain wall junction 
and how to decouple the unwanted mirror fermions.
In our formulation, there is an ambiguity in the choice of two domain walls,
which is discussed in Sec.~\ref{sec:2ndDW}.
Finally we propose how to regularize our formulation on a lattice in Sec.~\ref{sec:lattice}.
Section~\ref{sec:summary} is devoted to a summary and discussion.
Appendices~\ref{app:gamma}--\ref{app:muDWdetails} are given for technical details of our analysis. 

\section{{\it Parity} and $U(1)_A$ anomalies and related zero modes}
\label{sec:parity}

We consider fermion determinants
on a 6-dimensional Euclidean torus.
We take the gamma matrices $\gamma_i (i=1\cdots6)$ 
to be Hermitian and to satisfy $\{\gamma_i,\gamma_j\}=2\delta_{ij}$.
The 6-dimensional Dirac operator is denoted by $D^{\rm 6D}=\sum_{i=1}^6 \gamma_i \nabla_i$,
where $\nabla_i$ is the covariant derivative of a gauge group $SU(N_c)$.
Since we are interested in 4-dimensional gauge theory, 
we simply take the 5th and 6th components of the gauge fields to be zero, {\it i.e.}
$A_5(x)=A_6(x)=0$.
Later we define the remaining four components of the gauge fields in the bulk
by a two-parameter family of the 4-dimensional gauge fields localized 
at the domain-wall junction, but in this section we only require
$A_\mu (\mu=1,2,3,4)$ to be symmetric under $x_5\to -x_5$.

Let us start with a ratio of determinants of a single Dirac fermion and a Pauli-Villars field,
\begin{eqnarray}
\exp(-W_{\rm periodic})\equiv \det \left(\frac{D^{\rm 6D}-M-i\mu\gamma_6\gamma_7}{D^{\rm 6D}+M+i\mu\gamma_6\gamma_7}\right),
\label{eq:detperiodic}
\end{eqnarray}
where $\gamma_7=i\prod_{i=1}^6\gamma_i$ is the chirality operator, $M$ is the mass,
and $i\mu\gamma_6\gamma_7$ describes the constant axial vector current in the 6th direction,
\begin{equation}
i\mu \bar{\psi}\gamma_6\gamma_7\psi,
\end{equation}
which is invariant under the $U(1)_A$ rotation,
\begin{eqnarray}
g^\theta\psi(x) = e^{i\theta \gamma_7}\psi(x),\;\;\;\bar{\psi}(x)g^\theta=\bar{\psi}(x)e^{i\theta \gamma_7},
\end{eqnarray}
where $\theta$ is an arbitrary parameter.
Note that the Pauli-Villars field has the opposite signs of the masses $M$ and $\mu$.
Here we assume that the boundary condition is periodic in every direction,
in order to discuss the anomalies in the bulk 6-dimensions.
In later sections, we introduce the domain-walls to study the anomalies
at the boundaries.

%
Next, we introduce a {\it parity} transformation in the 5th direction on the fermion fields:
\begin{eqnarray}
P'\psi(x) = i\gamma_5 R_5 \psi(x),\;\;\;\bar{\psi}(x)P'=iR_5\bar{\psi}(x)\gamma_5,
\end{eqnarray}
where $R_i$ denotes the reflection of the $i$th coordinate: $R_i f(x_i)=f(-x_i)$.
Note that this parity is different from the conventional parity:
\begin{eqnarray}
P\psi(x)=\gamma_1\prod_{i\neq 1}R_i\psi(x),\;\;\;\bar{\psi}(x)P=\prod_{i\neq 1}R_i\bar{\psi}(x)\gamma_1,
\end{eqnarray}
where we take $i=1$ to be the temporal direction.
The main difference is that $P^{'2}=-1$ while $P^2=1$.
It is known that $P$-invariance exists only in even dimensions,
while $P'$ is allowed in any dimensions.
The massless Dirac fermion action 
\begin{equation}
S_F=\int d^6 x \bar{\psi}(x)D^{\rm 6D}\psi(x),
\end{equation}
has both of $P$ and $P'$ symmetries,
but the mass terms $M\bar{\psi}(x)\psi(x)$ and $i\mu\bar{\psi}(x)\gamma_6\gamma_7\psi(x)$ violate the $P'$ symmetry.

As is well known in the literature \cite{Redlich:1983dv, Niemi:1983rq, AlvarezGaume:1984nf}, 
the $P'$ symmetry has an anomaly.
Because of the anti-commutation relation
$\{D^{\rm 6D},P'\}=0$, every eigenvalue $i\lambda$ 
of $D^{6D}$ has its complex-conjugate pair $-i\lambda$,
except for the zero modes.
Therefore, under $P'$, the massless fermion action is manifestly invariant,
while the zero mode's contribution to the fermion measure Jacobian
is not, since $P'$ flips its sign,
\begin{eqnarray}
D\bar{\psi_0}P'DP'\psi_0 = - D\bar{\psi_0}D\psi_0.
\end{eqnarray}
Note that those from nonzero modes always cancel with their complex conjugates.
Therefore, the $P'$ transformation counts the number of zero modes $\mathfrak{I}$.

Moreover, using $P'$ and the axial $U(1)_A$ rotation\footnote{
In Eq.~(\ref{eq:chiral}), we have not used a naive operation $\exp(i\gamma_7\pi/2) = i\gamma_7$,
since the (lattice) regularization should break this condition.
}, with the angle $\theta=\pi$
\begin{eqnarray}
\label{eq:chiral}
g^\pi\psi(x)=\exp(i\gamma_7\pi/2)\psi(x),\;\;\;\bar{\psi}(x)g^\pi=\bar{\psi}(x)\exp(i\gamma_7\pi/2),
\end{eqnarray}
we can decompose $\mathfrak{I}$ into two parts,
\begin{eqnarray}
\mathfrak{I}=\mathcal{P}+\mathcal{I},
\end{eqnarray}
where $\mathcal{P}$ denotes the conventional index \cite{Atiyah:1963zz} related to the $U(1)_A$ anomaly,
namely $n_+-n_-$ where $n_\pm$ denote the number of zero modes 
with chirality $\pm$. The other integer $\mathcal{I}$ counts {\it exotic} zero modes,
which possibly exist even when the $U(1)_A$ anomaly is absent\footnote{
In the $SU(2)$ theory, for example, fermions cannot have
nonzero  $n_+-n_-$ in 6-dimensions since the $U(1)_A$ anomaly is zero.
Nevertheless, there exists the so-called mod-two index 
related to the homotopy group $\pi_5(SU(2))=\mathbb{Z}_2$. 
}.
As shown below, $\mathcal{P}$ controls the perturbative gauge anomaly,
while $\mathcal{I}$ can be considered as the origin of global anomalies.

The fermion determinant Eq.~(\ref{eq:detperiodic}) precisely 
reproduces this decomposition since
\begin{eqnarray}
\exp(-W_{\rm periodic}) &=& 
\det \left(\frac{D^{\rm 6D}-M-i\mu\gamma_6\gamma_7}{D^{\rm 6D}+M-i\mu\gamma_6\gamma_7}\right)\times
\det \left(\frac{D^{\rm 6D}+M-i\mu\gamma_6\gamma_7}{D^{\rm 6D}+M+i\mu\gamma_6\gamma_7}\right)
\nonumber\\
&=& \det \left(\frac{g^\pi(D^{\rm 6D}-M-i\mu\gamma_6\gamma_7)g^\pi}{D^{\rm 6D}+M-i\mu\gamma_6\gamma_7}\right)
\nonumber\\&&\times
\det \left(\frac{(g^\pi)^\dagger P'(D^{\rm 6D}+M-i\mu\gamma_6\gamma_7)P'(g^\pi)^\dagger}{D^{\rm 6D}+M+i\mu\gamma_6\gamma_7}\right)
\nonumber\\
&=& (-1)^{\mathcal{P}}\times(-1)^{\mathcal{I}},
\end{eqnarray}
where we have operated $g^\pi$ rotation to the numerator of the former determinant,
and $P'(g^\pi)^\dagger$ to the latter.
Note again that the $\mu$ term is $U(1)_A$ invariant. 

We find that the above argument does not change by replacing the $\mu$ term with
\begin{equation}
i\mu \bar{\psi}\gamma_6\gamma_7 R_5R_6\psi,\;\;\mbox{or}\;\;i\mu \bar{\psi}\gamma_6\gamma_7 R_6\psi.
\end{equation}
However, the nonlocal reflection operators $R_5$ or $R_6$
can make an unexpected cancellation of the
{\it physical} phase which should be present 
in the 4-dimensional target theory\footnote{
We thank D.~B.~Kaplan for pointing out this problem
in our original version of this paper, which was mainly analyzed
with the operator $i\mu \bar{\psi}\gamma_6\gamma_7 R_5R_6\psi$.
}. Therefore, in the following analysis, we use the simple 
axial vector current background operator.

\section{Two domain walls and anomaly inflow}
\label{sec:formulation}
\subsection{Weyl fermion at the domain wall junction}
Let us now give domain wall profiles to the two mass terms with $M$ and $\mu$:
\begin{eqnarray}
\exp(-W_{\rm 2DW})\equiv \det \left(\frac{D^{\rm 6D}+M\epsilon(x_6)
+i\mu\epsilon(x_5)\gamma_6\gamma_7  }{D^{\rm 6D}+M+i\mu\gamma_6\gamma_7  }\right),
\label{eq:detDW}
\end{eqnarray}
where $\epsilon(x)=x/|x|$ denotes the sign function.
Since the fermion fields satisfy periodic boundary conditions,
there also exist anti-domain walls in the determinant.
Although the anti-domain walls do not appear in the expressions,
we always assume that they are there, and 
will explicitly write them whenever it is necessary.

Decomposing the Dirac operator as
\begin{eqnarray}
D^{\rm 6D} &=& D^{\rm 4D} + \gamma_5\partial_5 + \gamma_6\partial_6,
\end{eqnarray}
where we have set $A_5=A_6=0$, we can
obtain a solution of the Dirac equation
\begin{eqnarray}
\label{eq:4Dsolution}
(D^{\rm 6D}+M\epsilon(x_6)+i\mu\epsilon(x_5)\gamma_6\gamma_7  )\psi(x)=0,
\end{eqnarray}
localized at the domain wall junction $x_5=x_6=0$ as
\begin{eqnarray}
\psi(x)&=&e^{-M|x_6|}e^{-\mu|x_5|}\phi(\bar{x}),\\
D^{\rm 4D} \phi(\bar{x})&=&0,\\
\gamma_6\phi(\bar{x})&=&\phi(\bar{x}),\\
i\gamma_5\gamma_6\gamma_7 \phi(\bar{x})&=&\phi(\bar{x}),
\end{eqnarray}
where $\bar{x}=(x_1,x_2,x_3,x_4)$ and we have assumed $M>0$ and $\mu>0$.
Note that $\gamma_6$ commutes with $i\gamma_5\gamma_6\gamma_7$, and
the two constraints by these operators force
$\phi(\bar{x})$ to have positive chirality (see Appendix~\ref{app:gamma}).
Namely, a Weyl fermion with positive chirality appears at the domain wall junction.
The Weyl fermion with the opposite chirality can be realized by flipping the signs of $M$ and $\mu$,
which changes the boundary conditions to $\gamma_6\phi(\bar{x})=-\phi(\bar{x})$ and 
$i\gamma_5\gamma_6\gamma_7\phi(\bar{x})=-\phi(\bar{x})$.
As will be shown below, the appearance of the single Weyl fermion is not a coincidence,
but required to keep the gauge invariance of the total 6-dimensional theory.

The total determinant Eq.~(\ref{eq:detDW}) becomes complex
due to the sign function $\epsilon(x_5)$, which is odd under $P'$.
We will see below that this complex phase that we denote by $\phi^\text{total}$ 
is {\it almost} localized
at the 4-dimensional junction of the two domain walls,
when the fermion contents are anomaly-free.


\subsection{Anomaly inflow through the $M$ domain wall}
In order to obtain the anomaly inflow through the $M$ domain wall,
first we consider a simpler case with $\mu=0$ and decompose the 
determinant into three parts,
\begin{eqnarray}
\label{eq:mu0limit}
\det \left(\frac{D^{\rm 6D}+M\epsilon(x_6)}{D^{\rm 6D}+M}\right)
&=&   \det \left(\frac{D^{\rm 6D}+M+iM_2\gamma_6\gamma_7  }{D^{\rm 6D}+M}\right)\nonumber\\
&&
\times \det \left(\frac{D^{\rm 6D}+M\epsilon(x_6)+iM_2\gamma_6\gamma_7  }{D^{\rm 6D}+M+iM_2\gamma_6\gamma_7  }\right)
\nonumber\\
&&\times\det \left(\frac{D^{\rm 6D}+M\epsilon(x_6)}{D^{\rm 6D}+M\epsilon(x_6)+iM_2\gamma_6\gamma_7  }\right),
\end{eqnarray}
and take the $M\gg M_2\gg 0$ limit. In this limit, there is no doubt that
the first determinant of the right hand side (RHS) converges to unity.
It is also important to note that in this $\mu\to 0$ limit,
the total determinant is real thanks to the $\gamma_7$ Hermiticity,
and the complex phase can be written as $\pi \mathfrak{I}$.

From the second determinant, we obtain the axial $U(1)_A$ anomaly:
\begin{eqnarray}
\label{eq:U(1)anomaly}
&&{\rm Im}\ln \det \left(\frac{D^{\rm 6D}+M\epsilon(x_6)+iM_2\gamma_6\gamma_7 }{D^{\rm 6D}+M+iM_2\gamma_6\gamma_7 }\right)\nonumber\\
&&={\rm Im}\ln \det \left(\frac{e^{i\theta(x_6)\gamma_7}e^{-i\theta(x_6)\gamma_7}(D^{\rm 6D}+M\epsilon(x_6)+iM_2\gamma_6\gamma_7 )
e^{-i\theta(x_6)\gamma^{reg}_7}e^{i\theta(x_6)\gamma^{reg}_7}}{D^{\rm 6D}+M+iM_2\gamma_6\gamma_7 }\right)\nonumber\\
&&= {\rm Im}\ln \det \left(e^{i\theta(x_6)\gamma_7}e^{i\theta(x_6)\gamma^{reg}_7}\right)\nonumber\\
&&= \pi\int d^6 x \frac{1-\epsilon(x_6)}{2}\frac{1}{6(4\pi)^3}
\epsilon^{\mu_1\cdots\mu_6}{\rm tr}[F_{\mu_1\mu_2}F_{\mu_3\mu_4}F_{\mu_5\mu_6}],
\end{eqnarray}
where $\theta(x_6)=\pi(1-\epsilon(x_6))/4$, and $\gamma_7^{reg}$ is the regularized chiral operator, 
for example, with the heat-kernel method, and the standard Fujikawa's method \cite{Fujikawa:1979ay} is applied.
Here, the $M_2\gg 0$ limit removes the IR divergence coming from the massless boundary-localized modes.
Since the integral in Eq.~(\ref{eq:U(1)anomaly}) counts the bulk instanton number 
in the region $x_6<0$, and gives a surface term at $x_6=0$, 
it can be decomposed as
\begin{eqnarray}
\pi \mathcal{P}^{6D}_{x_6<0}+\pi CS,
\end{eqnarray}
where $\mathcal{P}^{6D}_{x_6<0}$ is an integer
and $CS$ is the Chern-Simons term on the $M$ domain wall,
\begin{eqnarray}
CS&\equiv& -\int_{x_6=0} d^5 x \frac{2}{3(4\pi)^3}
\epsilon^{\mu_1\cdots\mu_5}{\rm tr}\left[\frac{1}{2}A_{\mu_1}F_{\mu_2\mu_3}F_{\mu_4\mu_5}
\right.\nonumber\\&&\left.
-\frac{i}{2}A_{\mu_1}A_{\mu_2}A_{\mu_3}F_{\mu_4\mu_5}
-\frac{1}{5}A_{\mu_1}A_{\mu_2}A_{\mu_3}A_{\mu_4}A_{\mu_5}
\right].
\end{eqnarray}

In the third determinant of Eq.(\ref{eq:mu0limit}), only the boundary localized mode 
at the $M$ domain wall can contribute. 
Projecting the determinant to the one for the sub-space which requires $\gamma_6\psi=\psi$,
and $(\gamma_6\partial_6+M\epsilon(x_6))\psi=0$, and rearranging the gamma-matrices, one obtains
\begin{equation}
\lim_{M\to \infty}\det \left(\frac{D^{\rm 6D}+M\epsilon(x_6)}{D^{\rm 6D}+M\epsilon(x_6)+iM_2\gamma_6\gamma_7 }\right)
= \det \left(\frac{\bar{D}^{\rm 5D}}{\bar{D}^{\rm 5D}+M_2}\right),
\label{eq:5Ddet}
\end{equation}
where the determinant in the RHS is taken in the reduced space 
of  4$\times$4 gamma matrices $\bar{\gamma}_i$ (see our notations in Appendix~\ref{app:gamma}), and 
the corresponding Dirac operator is given by
 $\bar{D}^{\rm 5D}=\sum_{i=1}^5\bar{\gamma}_i'\nabla_i|_{x_6=0}$.

Let us denote the complex phase of the determinant Eq.~(\ref{eq:5Ddet}) by
\begin{equation}
- i\pi\frac{\eta^{\rm 5D}}{2} = i{\rm Im}\det \left(\frac{\bar{D}^{\rm 5D}}{\bar{D}^{\rm 5D}+M_2}\right),
\label{eq:etadef}
\end{equation}
since $\eta^{\rm 5D}$ corresponds to a regularization of the so-called 
$\eta$-invariant \cite{AlvarezGaume:1985di, DellaPietra:1986qd, Kaplan:1995pe}:
\begin{equation}
\lim_{M_2\to\infty}\eta^{\rm 5D} = \sum_{\lambda>0}-\sum_{\lambda<0},  
\end{equation}
where $\lambda$ denote the eigenvalues of $i\bar{D}^{\rm 5D}$.
Therefore, we have obtained a formula
\begin{eqnarray}
\mathfrak{I} = \mathcal{P}^{6D}_{x_6<0}+CS - \frac{\eta^{\rm 5D}}{2},
\end{eqnarray}
known as the Atiyah-Patodi-Singer index theorem \cite{Atiyah:1975jf,Atiyah:1976jg,Atiyah:1980jh}.

Next, we turn on the $\mu$ domain wall and consider the limit $M\gg \mu\gg 0$.
A similar decomposition to Eq.~(\ref{eq:mu0limit}) is possible:
\begin{eqnarray}
\label{eq:MDW}
\det \left(\frac{D^{\rm 6D}+M\epsilon(x_6)+i\mu\epsilon(x_5) \gamma_6\gamma_7  }{D^{\rm 6D}+M+i\mu\gamma_6\gamma_7  }\right)
&=&   
\det \left(\frac{D^{\rm 6D}+M\epsilon(x_6)+i\mu\gamma_6\gamma_7 }{D^{\rm 6D}+M+i\mu\gamma_6\gamma_7 }\right)
\nonumber\\
&&\hspace{-0.25in}\times\det \left(\frac{D^{\rm 6D}+M\epsilon(x_6)+i\mu\epsilon(x_5)\gamma_6\gamma_7 }
{D^{\rm 6D}+M\epsilon(x_6)+i\mu\gamma_6\gamma_7 }\right).
\end{eqnarray}
The first determinant in Eq.(\ref{eq:MDW}) gives the same contribution as 
the product of first and second ones in Eq.~(\ref{eq:mu0limit}),
{\it i.e.} they produce the same $\pi(\mathcal{P}^{6D}_{x_6<0}+CS)$.
This is consistent with the fact that the chiral anomaly term is insensitive 
to the $\mu$ domain wall, which is $U(1)_A$ invariant.


The second determinant in Eq.(\ref{eq:MDW}) in the $M\to \infty$ limit, becomes
\begin{eqnarray}
\det\left(\frac{\bar{D}^{\rm 5D}+\mu\epsilon(x_5)}{\bar{D}^{\rm 5D}+\mu}\right),
\label{eq:det5DonMDW}
\end{eqnarray}
which needs a further decomposition into 
5-dimensional bulk and 4-dimensional boundary contributions.
While our target is the chiral fermion at the 4-dimensional junction at $x_5=0$,
the standard Pauli-Villars regulator requires the opposite chiral mode as well, 
to construct a mass term.
To this end, we explicitly write the anti-domain wall at $x_5=L_5$,
as was mentioned at the beginning of this section,
and introduce a nonlocal coupling to the fermion there.
More explicitly, we have
\begin{eqnarray}
\det\left(\frac{\bar{D}^{\rm 5D}+\mu\epsilon(x_5)\epsilon(L_5-x_5)}{\bar{D}^{\rm 5D}+\mu}\right)&=&
{\rm Det}\left(\frac{\delta(x-x')(\bar{D}^{\rm 5D}+\mu\epsilon(x_5)\epsilon(L_5-x_5))+\mu_2^{x_5,x_5'}}{\delta(x-x')(\bar{D}^{\rm 5D}+\mu)}\right)
\nonumber\\&&\hspace{-0.1in}
\times{\rm Det}\left(\frac{\delta(x-x')(\bar{D}^{\rm 5D}+\mu\epsilon(x_5)\epsilon(L_5-x_5))}{\delta(x-x')(\bar{D}^{\rm 5D}+\mu\epsilon(x_5)\epsilon(L_5-x_5))+\mu_2^{x_5,x_5'}}\right),
\nonumber\\
\label{eq:bedecom}
\end{eqnarray}
where ${\rm Det}$ denotes the determinant in the {\it doubled} space-time, 
so that we can insert a nonlocal mass term 
\begin{equation}
\mu_2^{x_5,x_5'}\equiv \mu_2 \left[
\delta(x_5)\delta(x_5'-L_5)+\delta(x_5-L_5)\delta(x_5')
\right].
\end{equation}
Note that this mass term violates the 5-dimensional gauge symmetry at $x_5=0$ and $x_5=L_5$ boundaries.
This term removes the contribution from the edge-localized modes in the first determinant of Eq.~(\ref{eq:bedecom}), 
while it plays a role of the UV cut-off in the second determinant,
which represents our target Weyl fermion.
In Appendix~\ref{app:bulkedge}, we present the details of this bulk/edge decomposition.

From the first determinant in Eq.~(\ref{eq:bedecom}), we obtain in its imaginary part another $CS$ term 
restricted to the $x_5<0$ region \cite{Callan:1984sa}: 
\begin{eqnarray}
-\pi CS^{(x_5<0)} &\equiv& \pi\int_{x_6=0} d^5 x \frac{4}{3(4\pi)^3}\frac{1-\epsilon(x_5)\epsilon(L_5-x_5)}{2}
\epsilon^{\mu_1\cdots\mu_5}{\rm tr}\left[\frac{1}{2}A_{\mu_1}F_{\mu_2\mu_3}F_{\mu_4\mu_5}
\right.\nonumber\\&&\left.
-\frac{i}{2}A_{\mu_1}A_{\mu_2}A_{\mu_3}F_{\mu_4\mu_5}
-\frac{1}{5}A_{\mu_1}A_{\mu_2}A_{\mu_3}A_{\mu_4}A_{\mu_5}
\right].
\end{eqnarray}
The gauge invariance is violated at the boundaries
in $CS^{(x_5<0)}$ due to the gauge noninvariant IR cutoff of the  boundary modes:
\begin{eqnarray}
-\pi \delta CS^{(x_5<0)} &=& -\frac{i}{24\pi^2} \int_{x_6=0, x_5=0} d^4 x 
\epsilon^{\mu_1\cdots\mu_4}{\rm Tr}\;v(x) \partial_{\mu_1}
\left[A_{\mu_2}\partial_{\mu_3}A_{\mu_4}+\frac{i}{2}A_{\mu_2}A_{\mu_3}A_{\mu_4}
\right](x)\nonumber\\
&&\hspace{-0.3in}+\frac{i}{24\pi^2} \int_{x_6=0, x_5=L_5} d^4 x 
\epsilon^{\mu_1\cdots\mu_4}{\rm Tr}\;v(x) \partial_{\mu_1}
\left[A_{\mu_2}\partial_{\mu_3}A_{\mu_4}+\frac{i}{2}A_{\mu_2}A_{\mu_3}A_{\mu_4}
\right](x),\nonumber\\
\label{eq:consistentanomaly}
\end{eqnarray}
where the gauge transformation is performed as $\delta A_\mu = -i\nabla_\mu v(x)$.
This form  exactly cancels the consistent anomaly \cite{Wess:1971yu, Bardeen:1984pm} 
of the Weyl fermions localized at $x_5=0$ and $x_5=L_5$,
and their cancellation is essential to keep the overall gauge invariance of the theory.

Before taking the $M\gg \mu$ limit, 
the phase of the second determinant in Eq.(\ref{eq:MDW}) may receive contributions from
exotic instantons, which are located in the region $x_5<0$.
In the limit of $M\gg \mu$, if these instantons are condensated
on the 5-dimensional $x_6=0$ plane,  they could give an integer contribution $\pi\mathcal{I}^\text{5D}_{M\gg\mu}$.

To the second determinant in Eq.~(\ref{eq:bedecom}), only the boundary Weyl fermion modes with positive chirality at $x_5=0$
and negative chirality at $x_5=L_5$ contribute, so that
\begin{eqnarray}
\lim_{\mu\to\infty}
{\rm Det}\left(\frac{\delta(x-x')(\bar{D}^{\rm 5D}+\mu\epsilon(x_5)\epsilon(L_5-x_5))}{\delta(x-x')(\bar{D}^{\rm 5D}+\mu\epsilon(x_5)\epsilon(L_5-x_5))+\mu_2^{x_5,x_5'}}\right)
= \det\frac{\mathcal{D}}{\mathcal{D}+\mu_2}
\label{eq:4ddet}
\end{eqnarray}
holds, where $\mathcal{D}$ is defined as
\begin{eqnarray}
\mathcal{D} = P^5_-\bar{D}^{\rm 4D}P^5_+ + P^5_+\bar{\partial}^{\rm 4D}P^5_-,
\end{eqnarray}
with $\bar{D}^{\rm 4D}=\sum_{i=1}^4\bar{\gamma}'_i\nabla_i|_{x_6=x_5=0}$ and $P^5_\pm=(1\pm \bar{\gamma}_5)/2$.
This determinant is not chiral gauge invariant and produces the consistent gauge anomaly
which is precisely canceled by Eq.~(\ref{eq:consistentanomaly}).
As will be shown later, we define the bulk gauge fields from  the 4-dimensional gauge fields
at the junction in such a way that the Dirac operator
$\bar{\partial}^{\rm 4D}=\sum_{i=1}^4\bar{\gamma}'_i\nabla_i|_{x_6=0,x_5=L_5}$ becomes 
that for a (almost) free fermion, so that the negative chiral mode
at $x_5=L_5$ is decoupled from the theory.

The complex phase of the determinant Eq.~(\ref{eq:4ddet}) is thus expressed by
\begin{eqnarray}
-i\frac{\pi}{2}\eta^{\rm 4D}+i\phi^{anom},
\end{eqnarray}
where $\eta^{\rm 4D}$ is the gauge invariant part, 
while $\phi^{anom}$ is the anomalous part\footnote{
If we can nonperturbatively evaluate 
the 4-dimensional determinant phase, 
and perform a random gauge transformation
on it, it is possible to smear out $\phi^{anom}$
and extract the gauge invariant phase $\eta^{\rm 4D}$.
It would be, however, practically difficult
to separate $\eta^{\rm 4D}$ and $\phi^{anom}$ of a single Weyl fermion.
Only the summation of $\eta^{\rm 4D}$ for the anomaly-free combination
would be possible to extract.
}.

What we have obtained is the anomaly ladder 
\begin{eqnarray}
\label{eq:ladderMDW}
\phi^\text{total}/\pi
&=& \mathcal{P}^{6D}_{x_6<0}+CS - \frac{\eta^{\rm 5D}}{2},\nonumber\\
\frac{1}{2}\eta^{\rm 5D}&=& CS^{(x_5<0)} + \mathcal{I}^\text{5D}_{M\gg\mu}+ \frac{1}{2}\eta^{\rm 4D}-\frac{\phi^{anom}}{\pi},
\end{eqnarray}
where $\mathcal{P}^{6D}_{x_6<0}$ denotes the 6-dimensional $U(1)_A$ anomaly,
$CS$ and $CS^{(x_5<0)}$ represent the 5-dimensional parity anomaly,
and $\phi^{anom}$ is the source for the consistent gauge anomaly.
This result is consistent with the anomaly descent equations 
found by Stora \cite{Stora:1983ct} 
and Zumino \cite{Zumino:1983ew, Zumino:1983rz},
including the overall constant determined by 
Alvarez-Gaum\'e and Ginsparg \cite{AlvarezGaume:1983cs} 
and Sumitani \cite{Sumitani:1984ed}.

When the (perturbative) gauge anomaly is absent, as well as
the number of fundamental fermions is even (to cancel the global anomaly)
the formula reduces to
\begin{equation}
\label{eq:totalphaseM}
\sum_f\phi_f^\text{total}=-\sum_f\frac{\pi}{2}\eta_f^{\rm 4D},
\end{equation}
mod $2\pi$, where we have put the subscript $f$ to represent the flavor of different fermions.
This means that the complex phase of the total fermion determinant
essentially comes from the 4-dimensional edge modes, 
at least in the hierarchical limit of $M\gg \mu\gg 0$.


\subsection{Anomaly inflow through the $\mu$ domain wall}

In order to trace the anomaly inflow via the $\mu$ domain wall,
let us take the limit $\mu\gg M\gg 0$.
Ignoring $M$, the fermion determinant can be decomposed into three parts,
\begin{eqnarray}
\label{eq:M0limit}
\det \left(\frac{D^{\rm 6D}+i\mu\epsilon(x_5)\gamma_6\gamma_7 }{D^{\rm 6D}+i\mu\gamma_6\gamma_7 }\right)
&=&  \det \left(\frac{D^{\rm 6D}+i\mu\gamma_6\gamma_7 +\mu_2}{D^{\rm 6D}+i\mu\gamma_6\gamma_7 }\right)
\nonumber\\
&&
\times \det \left(\frac{D^{\rm 6D}+i\mu\epsilon(x_5)\gamma_6\gamma_7 +\mu_2}{D^{\rm 6D}+i\mu\gamma_6\gamma_7 +\mu_2}\right)
\nonumber\\
&&\times\det \left(\frac{D^{\rm 6D}+i\mu\epsilon(x_5)\gamma_6\gamma_7 }{D^{\rm 6D}+i\mu\epsilon(x_5)\gamma_6\gamma_7 +\mu_2}\right),
\end{eqnarray}
where we take the $\mu\gg \mu_2\gg 0$ limit. 
Note that the first determinant converges to unity.

Unlike the $M$ domain wall, the second determinant does not produce the axial $U(1)$ anomaly. 
Due to the explicit violation of the $SO(6)$ Lorentz symmetry by the axial vector background,
the phase $\phi^\text{6D}$ of the second determinant can be expanded in an $SO(5)$ invariant series of $1/\mu$,
except for the nonperturbative zero mode's contribution 
$\pi \mathcal{I}^{\rm 6D}_{x_5<0}$, which is located in the region $x_5<0$.
More explicitly, we have
\begin{eqnarray}
\phi^\text{6D} = \pi \mathcal{I}^{\rm 6D}_{x_5<0} + \mu \phi^{(1)} + \phi^{(2)}/\mu +\mathcal{O}(1/\mu^3),
\end{eqnarray}
where, the leading order contribution has a form of the Chern-Simons term
\begin{eqnarray}
 \phi^{(1)}=c_0\pi\int d^6 x \frac{4}{3(4\pi)^3}\frac{1-\epsilon(x_5)}{2}
\epsilon^{i_1\cdots i_5}{\rm tr}\left[\frac{1}{2}A_{i_1}F_{i_2 i_3}F_{i_4 i_5}
\right.\nonumber\\\left.
-\frac{i}{2}A_{i_1}A_{i_2}A_{i_3}F_{i_4 i_5}
-\frac{1}{5}A_{i_1}A_{i_2}A_{i_3}A_{i_4}A_{i_5}
\right],
\end{eqnarray}
and the next-to-leading order contribution is 
\begin{eqnarray}
 \phi^{(2)}=\int d^6 x
\frac{1-\epsilon(x_5)}{2} 
\epsilon^{i_1\cdots i_5}
{\rm tr}\left[c_1 D^j F_{i_1j}F_{i_2 i_3}F_{i_4 i_5}
+c_2 D^{6} F_{6j}F_{i_2 i_3}F_{i_4 i_5}
\right.\nonumber\\\left.
+c_3 D_{i_1} F_{6 i_2}F_{6 i_3}F_{i_4 i_5}
\right],
\end{eqnarray}
where $c_k$ are numerical constants\footnote{
We find that $c_0$ is logarithmically divergent since a single 
Pauli-Villars spinor is not enough to regularize the determinant.
Modifying the second determinant of Eq.~(\ref{eq:M0limit}) to 
$\det \left(\frac{D^{\rm 6D}+i\mu\epsilon(x_5)\gamma_6\gamma_7 +\mu_2}{D^{\rm 6D}+i\mu\gamma_6\gamma_7 +\mu_2}\right)\times \det\left(\frac{D^{\rm 6D}+i\Lambda\gamma_6\gamma_7 +\mu_2}{D^{\rm 6D}+i\Lambda\epsilon(x_5)\gamma_6\gamma_7 +\mu_2}\right)^{\frac{\mu}{\Lambda}}$, 
we obtain a finite value of $c_0$ proportional to $\ln \Lambda$. 
Here, $\Lambda\gg \mu\gg \mu_2\gg 0$ is assumed.
} determined by the representation of the fermion,
and the summation of the indices is taken in the 5-dimensions only.
In the above computation, $\mu_2$ plays a role of an infra-red cut-off,
removing the contribution from the edge-localized modes.

As will be discussed later, when the theory is free  from 
both perturbative and global anomalies, 
the total complex phase of the determinants is
\begin{eqnarray}
\sum_f\phi_f^\text{6D} =  \sum_f \phi_f^{(2)}/\mu_f+\mathcal{O}(1/\mu_f^3),
\end{eqnarray}
which disappears in the limit $\mu_f\to \infty$.
Here the subscript $f$ denotes the flavor index.

In the third determinant in Eq.~(\ref{eq:M0limit}), only the boundary localized modes 
on the $\mu$ domain wall, which satisfy $i\gamma_5\gamma_6\gamma_7 \psi =\psi$ can contribute. 
Here we see a significant difference from the previous subsection, where
we obtained a single massless Dirac fermion determinant in Eq.~(\ref{eq:5Ddet}).
What we obtain here is not
a single Dirac fermion but two (4-component) Dirac fermions having Pauli-Villars masses $\pm \mu_2$ with opposite signs, 
that are constrained to have the positive eigenvalue of the gamma matrix $\bar{\gamma}_5$:
\begin{eqnarray}
&&\lim_{\mu \to \infty}\det \left(\frac{D^{\rm 6D}+i\mu\epsilon(x_5)\gamma_6\gamma_7 }
{D^{\rm 6D}+i\mu\epsilon(x_5)\gamma_6\gamma_7 +\mu_2}\right)
\nonumber\\&&
= \det \left(P_+\frac{\hat{D}^{\rm 5D}}{\hat{D}^{\rm 5D}+\mu_2}P_+  +P_-\right)
\det \left(P_+\frac{\hat{D}^{\rm 5D}}{\hat{D}^{\rm 5D}-\mu_2}P_+  +P_-\right),
\label{eq:5Ddetmu}
\end{eqnarray}
where $\hat{D}^{\rm 5D}=(\sum_{i=1}^4\bar{\gamma}_i \nabla_i+\bar{\gamma_5}\partial_6)|_{x_5=0}$,
and $P_\pm \equiv (1\pm\bar{\gamma}_5)/2$ is a Hermitian projection operator.
In Appendix~\ref{app:muDWdetails} we present the details of our computation.

The determinant Eq.~(\ref{eq:5Ddetmu}) is real and
its phase can be defined as $\pi \mathcal{I}^{\rm 5D}$\footnote{
The formula looks not only real but positive. 
The nontrivial phase $\pi \mathcal{I}^{\rm 5D}$, however, comes from the zero-modes, 
where the determinant becomes ill-defined.
Therefore, in order to precisely compute $\pi \mathcal{I}^{\rm 5D}$, we need
a careful massless limit from the massive determinant, as well as
an appropriate regularization 
to count the number of exotic zero-modes $\mathcal{I}^{\rm 5D}$.
Since a good regularization should not break the complex conjugate pairs of nonzero modes,
we can generally claim that the phase is $\pi \mathcal{I}^{\rm 5D}$.
}.
This means that only the mod-two type exotic index
can communicate with the lower-dimensions.
There appears no CS term,
and therefore, no source of the perturbative gauge anomaly 
on the $\mu$ domain wall.

Now, let us turn on the $M$ domain wall and consider the limit $\mu\gg M\gg 0$:
\begin{eqnarray}
\label{eq:muDW}
\det \left(\frac{D^{\rm 6D}+M\epsilon(x_6)+i\mu\epsilon(x_5) \gamma_6\gamma_7  }{D^{\rm 6D}+M+i\mu\gamma_6\gamma_7  }\right)
&=& 
\det \left(\frac{D^{\rm 6D}+i\mu\epsilon(x_5)\gamma_6\gamma_7 +M}{D^{\rm 6D}+i\mu\gamma_6\gamma_7 +M}\right)
\nonumber\\
&&\times\det \left(\frac{D^{\rm 6D}+i\mu\epsilon(x_5)\gamma_6\gamma_7 +M\epsilon(x_6)}{D^{\rm 6D}+i\mu\epsilon(x_5)\gamma_6\gamma_7 +M}\right),
\nonumber\\ 
\end{eqnarray}
where the first determinant gives the same result as those in Eq.~(\ref{eq:M0limit}).
Namely they produce the phase $\phi^\text{6D}$.


The second determinant in Eq.~(\ref{eq:muDW}) in the $\mu\to \infty$ limit is\footnote{
As $\hat{D}^{\rm 5D}$ and $P_+$ do not commute with $M\epsilon(x_6)$, we make the order of
the matrix operations explicit.} 
\begin{eqnarray}
\label{eq:4thdetmuDW}
&&\det \left(P_+(\hat{D}^{\rm 5D}+M)^{-1}(\hat{D}^{\rm 5D}+M\epsilon(x_6))P_+ +P_-\right)
\nonumber\\&&
\times\det \left(P_+(\hat{D}^{\rm 5D}-M)^{-1}(\hat{D}^{\rm 5D}-M\epsilon(x_6))P_+ +P_- \right).
\end{eqnarray}
This expression is {\it almost} real, except for the domain wall $x_6=0$,
since the complex phase comes from the noncommutativity of $\hat{D}^{\rm 5D}$
and $M\epsilon(x_6)$, which is proportional to $\delta(x_6)$.

Let us further decompose Eq.~(\ref{eq:4thdetmuDW}) as
\begin{eqnarray}
\label{eq:4thdetmuDW2}
&&\det \left(P_+(\hat{D}^{\rm 5D}+M)^{-1}(\hat{D}^{\rm 5D}+M\epsilon(x_6)+M_2)P_+ +P_- \right)
\nonumber\\&&
\times\det \left(P_+(\hat{D}^{\rm 5D}-M)^{-1}(\hat{D}^{\rm 5D}-M\epsilon(x_6)-M_2)P_+ +P_- \right)
\nonumber\\&&
\times\det \left(P_+(\hat{D}^{\rm 5D}+M\epsilon(x_6)+M_2)^{-1}(\hat{D}^{\rm 5D}+M\epsilon(x_6))P_+ +P_- \right)
\nonumber\\&&
\times\det \left(P_+(\hat{D}^{\rm 5D}-M\epsilon(x_6)-M_2)^{-1}(\hat{D}^{\rm 5D}-M\epsilon(x_6))P_+ +P_- \right)
\end{eqnarray}
and take the limit $M\gg M_2\gg 0$. In this case, $M_2$ cannot completely separate
the bulk and the edge modes, due to the projection operator $P_+$.
Here we can only say that the total determinant is complex, 
whose phase is almost localized at $x_6=0$.

In the third and fourth determinants in Eq.~(\ref{eq:4thdetmuDW2}),
we observe an interesting dynamics.
First of all, this combination of two determinants is real and positive.
Therefore, the nontrivial complex phase resides in the first and 
second determinants, in a gauge invariant way. 
Secondly, the Weyl fermions appear only in the third determinant, since
only the positive chirality mode survives the projection $P_+$.
The fourth determinant then contains the contribution from the bulk modes 
(and possibly from the doubler modes when we take a lattice regularization).
Therefore, we are left with
\begin{eqnarray}
&&\lim_{M\to\infty}\det \left(P_+(\hat{D}^{\rm 5D}+M\epsilon(x_6)+M_2)^{-1}
(\hat{D}^{\rm 5D}+M\epsilon(x_6))P_+ + P_-\right)
\nonumber\\&&\times
\det \left(P_+(\hat{D}^{\rm 5D}-M\epsilon(x_6)-M_2)^{-1}(\hat{D}^{\rm 5D}-M\epsilon(x_6))P_+ +P_- \right)
\nonumber\\
&\propto& \det\left(P^5_+(\bar{D}^{\rm 4D}+M_2)^{-1}\bar{D}^{\rm 4D}P^5_++P_-\right)\times\exp(-i\phi^{nl})
\label{eq:4ddet2}
\end{eqnarray}
where $\bar{D}^{\rm 4D}=\sum_{i=1}^4\bar{\gamma}_i\nabla_i|_{x_6=x_5=0}$, 
representing a single Weyl fermion determinant.
Note that the phase $\phi^{nl}$ cannot be written as any local effective action in 4-dimensions.
However, we already know its origin. It is the CS term on the $M$ domain wall.
It is hidden in the nonlocal phase  $\phi^{nl}$ since we have integrated the bulk contribution
in the 5-th direction first.

From the above analysis, we may write 
the phase of the second determinant in Eq.~(\ref{eq:muDW}) as 
\begin{eqnarray}
\pi \mathcal{I}^{\rm 5D}_{x_6<0}-\frac{\pi}{2}\eta^{\rm 4D}, 
\end{eqnarray}
up to some regularization dependent term (which will be neglected below)\footnote{
Since the $M\gg \mu\gg 0$ and $\mu \gg M\gg 0$ limits may give
different regularizations of the low-energy effective theory,
the remaining phase in the 4-dimensional junction may differ 
by $f(\infty)-f(0)$, where $f(M/\mu)$ is the regularization dependent function.
}.
Then the phase of the total 6D determinant can be decomposed as
\begin{eqnarray}
\phi^\text{total} = \pi \mathcal{I}^{\rm 6D}_{x_5<0} + \mu \phi^{(1)} + \phi^{(2)}/\mu + 
\pi \mathcal{I}^{\rm 5D}_{x_6<0}-\frac{\pi}{2}\eta^{\rm 4D} +\mathcal{O}(1/\mu^3),
\end{eqnarray}
where $\mathcal{I}^{\rm 6D}_{x_5<0}$ and $\mathcal{I}^{\rm 5D}_{x_6<0}$ are
the exotic indices in the 6-dimensional bulk and 
the 5-dimensional $\mu$ domain wall, respectively.
When the theory has an anomaly-free combination of the fermion flavors, 
the total phase is
\begin{eqnarray}
\sum_f\phi_f^\text{total} = -\sum_f\frac{\pi}{2}\eta_f^{\rm 4D} +\mathcal{O}(1/\mu_f),
\end{eqnarray}
where $f$ denotes the flavor index.
Namely, the complex phase is determined by the fermion modes localized at the
4-dimensional junction in the $\mu\to \infty$ limit, which is consistent with 
another $M\gg \mu\gg 0$ limit already seen in Eq.~(\ref{eq:totalphaseM}).

\subsection{domain wall junction}

In the previous two subsections, we have traced two different anomaly inflows taking $M\gg \mu \gg 0$ 
and $\mu \gg M \gg 0$ limits.
At finite $M$ and $\mu$, the situation can be more complicated but
the nontrivial cancellation of anomalies among different dimensions
should be maintained to keep the gauge invariance of the total theory.
In the end, a single Weyl fermion always appears at the junction of the two domain walls.

\begin{figure}[htbp]
\begin{center}
\includegraphics[width=10cm, angle=90]{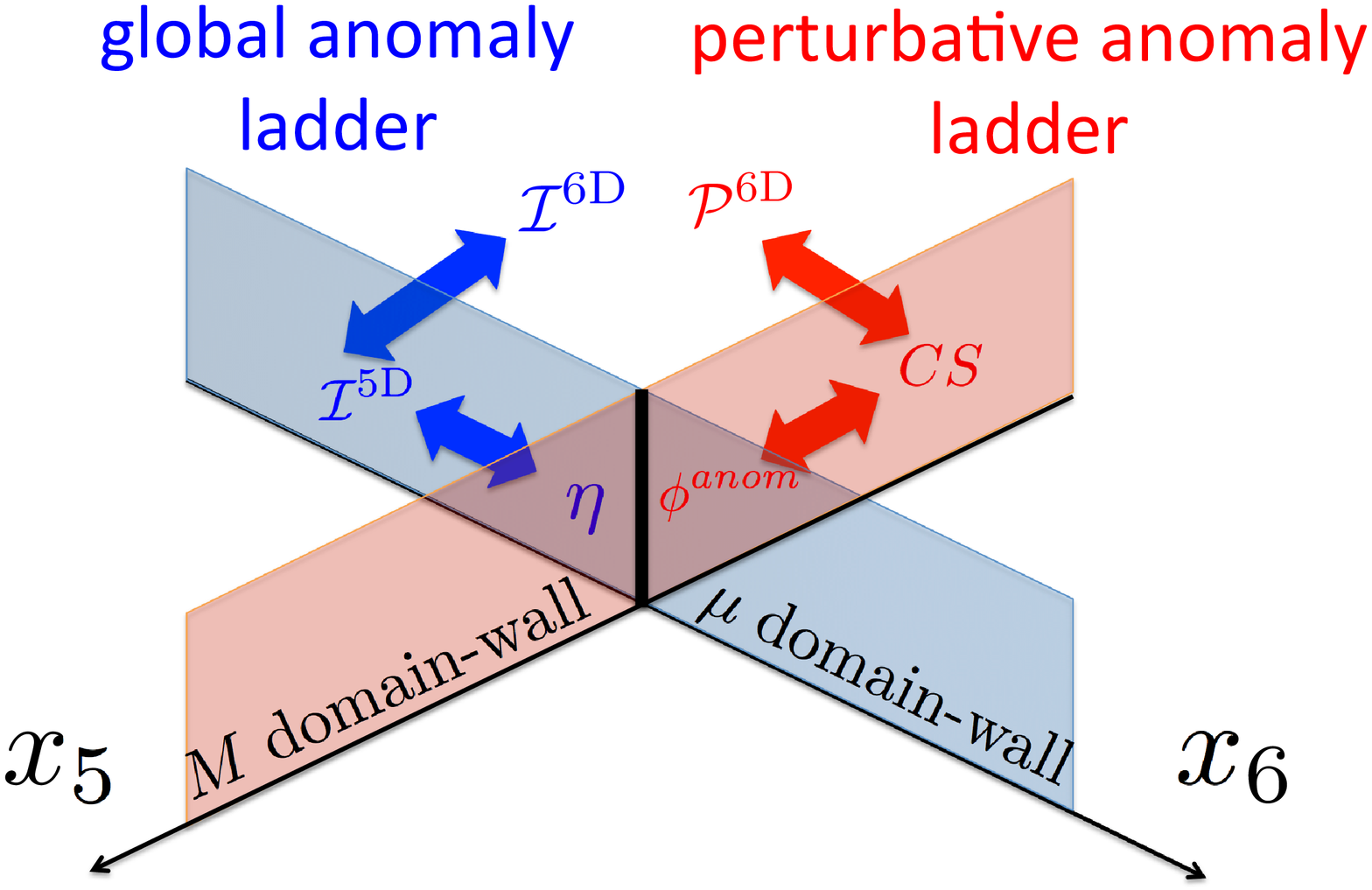}
\caption{The anomaly inflows through the two domain walls. 
The $M$ domain wall at $x_6=0$ 
mediates the perturbative anomaly inflow (red arrows), 
which exhibits the Stora-Zumino descent equations.
The $\mu$ domain wall at $x_5=0$ mediates 
the inflow of the global anomaly (blue). 
} 
\label{fig:2DW}
\end{center}
\end{figure}

When a small gauge transformation is performed at the 4-dimensional junction,
the gauge current flows through the $M$ domain wall,
but never flows into the $\mu$ domain wall, 
since there is no CS term which can absorb the gauge noninvariance.
Instead, a large gauge transformation can create exotic instantons on
the $\mu$ domain wall and flip the sign of the partition function. 
Thus, we confirm that the perturbative anomaly inflow, 
which naturally exhibits the Stora-Zumino anomaly descent equations,
is mediated by the $M$ domain wall, while the inflow of the global anomaly 
goes through the $\mu$ domain wall 
(see Fig.~\ref{fig:2DW}).



\section{Anomaly free condition}
\label{sec:anomalyfree}

Due to the topological obstructions of 
the $U(1)_A$ and $P'$ symmetries,
a single Weyl fermion cannot be described by 
a 4-dimensional local field theory.
If these anomalies are canceled among different flavors,
the net anomaly inflow down to the 4-dimensional junction vanishes,
and the chiral gauge current can be conserved,
realizing a consistent 4-dimensional theory in the low-energy limit.

The cancellation of the $U(1)_A$ anomaly is assured if
\begin{eqnarray}
\label{eq:pfree}
\sum_L {\rm tr}T_L^a\{T_L^b,T_L^c\}
-\sum_R {\rm tr}T_R^a\{T_R^b,T_R^c\}=0,
\end{eqnarray}
where $T_{L/R}$ denote the gauge group generators
in $L/R$ representation of the corresponding 
left/right handed fermions.
This is the well-known anomaly free condition
of the perturbative chiral gauge invariance.
In our formulation, the condition Eq.~(\ref{eq:pfree})
guarantees the cancellation of 
the $U(1)_A$ anomaly as well as the CS term
on the $M$ domain-wall, so that the gauge current
never flows out of the 4-dimensional junction.

For global anomalies in 4-dimensions, 
it is usual to consider only the case with $SU(2)$ group.
This is because the map from a 4-dimensional sphere $S^4$
to the gauge group $G$: $\pi_4(G)$, is only nontrivial for $SU(2)$.
In our formulation, this $SU(2)$ anomaly is 
embedded as the phase of the 6-dimensional Dirac fermions
through the APS(-like) index relation
\begin{eqnarray}
\pi_5(SU(2))=\mathbb{Z}_2 \sim \pi_4(SU(2)) =\mathbb{Z}_2.
\end{eqnarray}
This homomorphism is not found in the literature on physics.
To cancel the $SU(2)$ anomaly, we need even number of 
fundamental fermions so that the gauge transformation never
flips the sign of the total partition function.

However, the cancellation of the global anomalies 
is more nontrivial, as discussed in \cite{Dai:1994kq, Witten:2015aba}.
The global anomaly should be absent not 
only on a simple manifold like $S^4$ or $S^5$
but also on any compact manifold. 
Our setup on the 6-dimensional torus having
 domain wall junctions of 4-dimensional torus,
is already such a nontrivial example.

In fact, L\"uscher found in the construction of 
a $U(1)$ chiral gauge theory on the lattice \cite{Luscher:1998du},
that a condition 
\begin{eqnarray}
\mbox{number fermions with odd charges = even},
\end{eqnarray}
is required to keep the nonperturbative chiral gauge invariance,
although it was not clearly identified as one of global anomalies\footnote{
Similar inconsistencies of $U(1)$ chiral gauge theories on two-dimensional torus 
were also reported in  \cite{Narayanan:1996cu} and \cite{Izubuchi:1999km}.
It was also reported in Ref.~\cite{Bar:2002sa} that the 
global anomaly cancellation in the $SU(2)$ theory 
for $\frac{4n+3}{2}$ representations is not trivial.
The global anomalies for these cases on various manifold
may need to be re-examined on various manifolds.
}.
This is not surprising since on the 4-dimensional torus 
$T^4=S^1\times S^1\times S^1 \times S^1$, at least one cycle
may develop a nontrivial map: $\pi_1(U(1))=\mathbb{Z}$,
even when the perturbative anomaly is absent.

In this paper, we do not try to extensively classify
the global anomalies but just mention that if
\begin{eqnarray}
\label{eq:gfree}
\mbox{number fermions in the fundamental representation = even},
\end{eqnarray}
after the irreducible decomposition, 
our 6-dimensional theory is free from the global anomalies that originate
from the exotic index $\mathcal{I}$.
The standard model of particle physics satisfies the above condition if 
we identify $e/6$ as a unit charge of the hyper-charge.

The above anomaly free conditions are those which
must be satisfied in the continuum limit.
At a finite cut-off, we have to further control 
the remaining violation of the gauge invariance,
since the anomaly cancellation is not perfect.
This is due to the fact that the bulk determinant 
respects the 6-dimensional gauge invariance,
which is not the one in our target 4-dimensional theory.
As will be discussed in the next section 
we follow the strategy in Ref.~\cite{Grabowska:2015qpk}
to use the Yang-Mills gradient flow in 5-th and 6-th directions.
The gradient flow realizes a kind of {\it dimensional extension}
so that the fermions in the extra (flavor) space, 
share the same 4-dimensional gauge invariance.

One disadvantage of taking the gradient flow both in
5-th and 6-th directions is that the role of 
the $\mu$  domain wall to detect the global anomaly becomes obscure.
Since the flowed gauge fields are invariant under
any gauge transformations, 
it is unlikely to have nonzero index $\mathcal{I}$ on the 
$\mu$  domain wall, which requires a nontrivial response to
large gauge transformations.
This means that the lattice formulation cannot detect inconsistencies of the gauge theory with 
odd number of flavors which is anomalous under global gauge transformation. 
To circumvent this problem one should look for a better formulation which uses an extended 
gauge  field sensitive to the global anomaly yet keeping the perturbative gauge invariance.
We leave it as an open problem. 

In this work, we take the following practical solution which is similar 
in spirit to Ref.~\cite{Witten:2015aba}.
It is argued in Ref.~\cite{Witten:2015aba} that
some global anomalies cannot be detected on the mapping torus, 
which is a standard setup to discuss the global anomalies,
but can appear on other manifolds.
In such theories, the mapping torus is in a sense an {\it unlucky} 
setup which cannot distinguish the anomalous and nonanomalous fermion contents.
We may regard our setup using the Yang-Mills gradient flow as a similar unlucky example.
Namely, to discuss the both of the perturbative and global anomalies,
we should use general background of 6-dimensional gauge fields.
Once the anomaly free conditions are obtained in this general setup, 
then we may restrict the gauge fields using the gradient flow,
to construct the target 4-dimensional gauge theory.


\section{Decoupling of the mirror fermions}
\label{sec:4dtheory}

So far we have not discussed the effects of 
the anti-domain-walls.
In order to realize a single Weyl fermion,
the massless modes at other domain-wall junctions
must be decoupled from the theory.

First, we take the spatial extents in the 5-th and 6-th directions 
to be finite in the ranges $-L_5 < x_5\leq L_5$ and $-L_6 < x_6\leq L_6$.
We take the fermion fields to satisfy periodic boundary conditions,
which requires (at least) one $M$ anti-domain-wall at $x_6=L_6 (=-L_6)$
and one $\mu$ anti-domain-wall at $x_5=L_5 (=-L_5)$.
Our fermion determinant is now
\begin{eqnarray}
\exp(-W_{\rm 2DW})= \det \left(\frac{D^{\rm 6D}+M\epsilon(x_6)\epsilon(L_6-x_6)
+i\mu\epsilon(x_5)\epsilon(L_5-x_5)\gamma_6\gamma_7 }{D^{\rm 6D}+M+i\mu\gamma_6\gamma_7 }\right),
\label{eq:detDWfiniteV}
\end{eqnarray}
in which we have 4 domain-wall junctions.
Two Weyl fermion modes with positive chirality appear 
at $(x_5,x_6)=(0,0)$ and $(0,L_6)$, while
those with negative chirality are localized at
$(x_5,x_6)=(L_5,0)$ and $(L_5,L_6)$.

Among these 4 junctions, only the one at $(x_5,x_6)=(0,0)$
is needed to construct {\it our world} in 4-dimensions,
and we would like the Weyl fermions at other three junctions
to be decoupled from the gauge fields.
To achieve this, we use the profile of the gauge field
in the fifth and sixth directions using the Yang-Mills gradient flow,
following the idea in \cite{Grabowska:2015qpk}.
The gradient flow exponentially weaken the gauge fields with the flow time
so that the Weyl fermions at $x_5=L_5$ and $x_6=L_6$ are decoupled from the gauge fields.
As flowed gauge fields transform in the same way as the original fields,
we can maintain the 4-dimensional gauge invariance of the total theory. 

\begin{figure}[htbp]
\begin{center}
\includegraphics[width=10cm, angle=90]{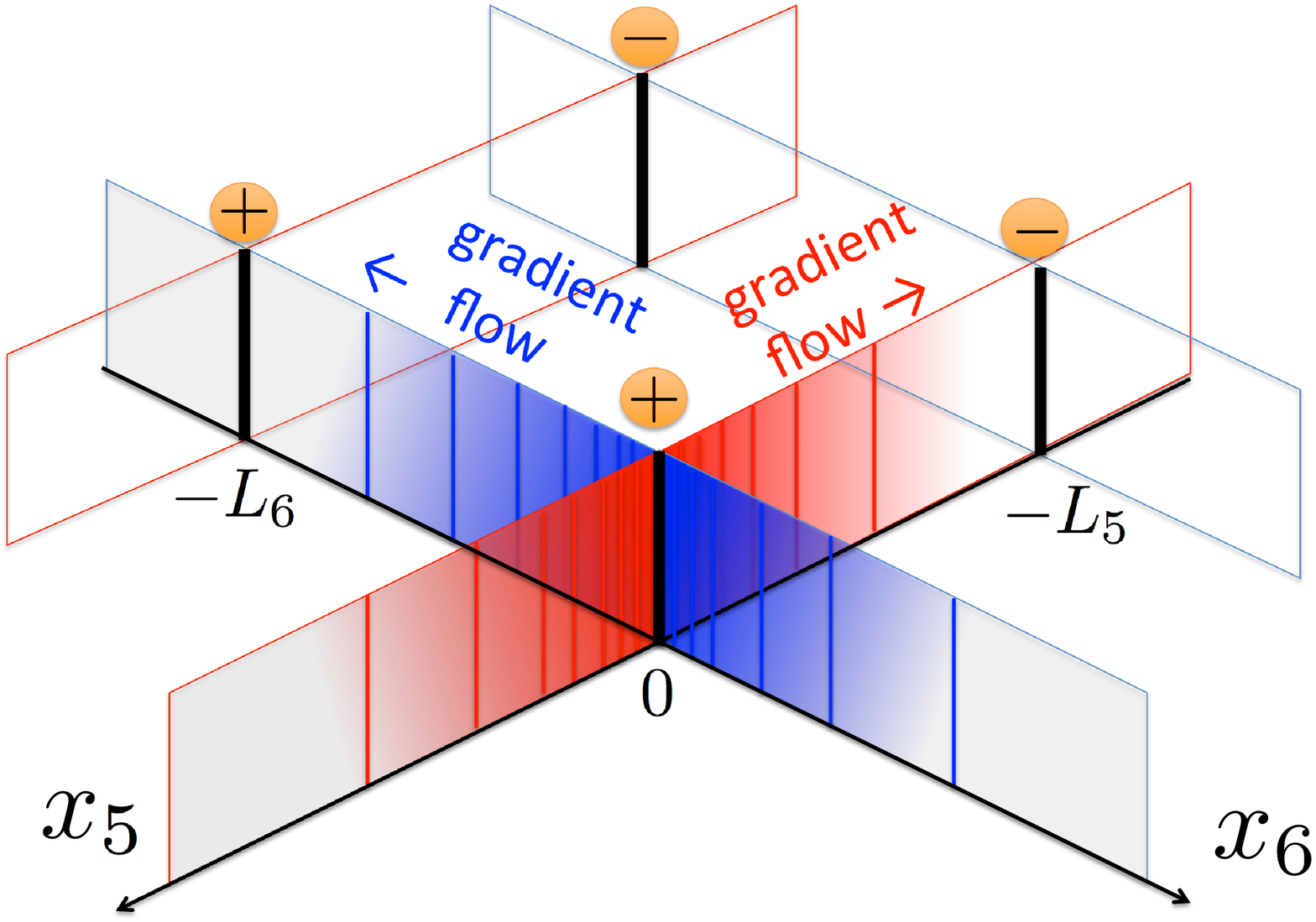}
\caption{Schematic view of our 6-dimensional finite space.
The $\pm$ symbols show the Weyl modes with positive and negative chiralities,
localized at each of the four domain-wall junctions.
The case with $M>0$ and $\mu>0$ is shown.
Our target Weyl fermion with positive chirality is localized at
the origin, while other three Weyl fermions are decoupled from
the gauge fields by the gradient flow.
} 
\label{fig:FiniteLattice}
\end{center}
\end{figure}

More explicitly, we take 
\begin{eqnarray}
A_\mu(\bar{x},x_5,x_6) =
A_\mu^{|x_5|+|x_6|}(\bar{x})\;\;\;\mbox{($\mu=1,\cdots 4$)},\;\;\;
A_5 = A_6 = 0,
\end{eqnarray}
where $\bar{x}=(x_1,x_2,x_3,x_4)$ is the coordinate of 
the 4-dimensional torus.
$A_\mu^{t}$ denotes the solution of the Yang-Mills gradient flow at a flow time $t$,
\begin{eqnarray}
\frac{\partial}{\partial t}A_\mu^t &=& \frac{\xi \epsilon(t)}{M}D_\nu F^t_{\nu\mu} \;\;\;\mbox{($\mu,\nu =1,\cdots 4$)}, 
\end{eqnarray}
where $D_\nu$ and $F^t_{\mu\nu}$ are the covariant derivatives and 
field strengths with respect to the flowed gauge field $A_\mu^t$, respectively.
$\xi$ is an arbitrary constant of order one.
Here $A_\mu^0 \equiv A_\mu(\bar{x})$ is 
the physical dynamical variable over which we integrate
in the path integral.
Our finite volume set-up is shown in Fig.~\ref{fig:FiniteLattice}.

Recently, Okumura and Suzuki \cite{Okumura:2016dsr} found that
the mirror fermions in the 4-dimensional effective theory 
\cite{Grabowska:2016bis,Makino:2016auf} using the Yang-Mills gradient flow
in 5-dimensional domain-wall set up
are not completely decoupled from the gauge fields.
This can be seen by the exact conservation of the total fermion 
numbers of physical and  mirror fermions, which implies
that the mirror fermions are sensitive to the topology of the
original gauge fields even after the gradient flow,
and the resulting theory should have non-local properties
due to this remnants of mirror fermions.

This problem of non-locality is inherited to our 6-dimensional model,
unless we give up employing the Yang-Mills gradient flow.
Since the procedure of fixing the 6-dimensional gauge fields
using the 4-dimensional configuration itself is already non-local
in terms of 6-dimensional quantum field theory,
it might be safer if we can achieve a mechanism of decoupling mirror fermions
in a local and dynamical way in the 6-dimensional field theory set-ups.
However, we have not found any such formulation realizing the
localization of gauge fields at the domain-wall junction.

\section{The choice of $\mu$ domain wall operator}
\label{sec:2ndDW}

In this work, we have chosen the axial vector back ground 
$i\mu \epsilon(x_5)\gamma_6\gamma_7$, which is insensitive to $U(1)_A$,
to realize the $\mu$ domain wall.
This choice is, however, not the unique solution for having chiral mode at the domain wall junction.
For example, we find that for the operators
\begin{eqnarray}
i\mu \epsilon(x_5)\gamma_6\gamma_7R_6,\;\;\mbox{or}\;\;i\mu \epsilon(x_5)\gamma_6\gamma_7 R_5 R_6,
\end{eqnarray}
the 4-dimensional localized solution in Eq.~(\ref{eq:4Dsolution}) is unchanged.
The structure of the anomalies is, however, different among these operators.
In particular, the use of $i\mu \epsilon(x_5)\gamma_6\gamma_7 R_5 R_6$ makes the
total fermion determinant real, even when the theory is anomalous.
It seems that the non-locality induced by the reflection operators $R_5$ and $R_6$
makes an unwanted cancellation of the complex phase, 
including the phase that should survive in the continuum limit.

It is unclear if the $\mu$ domain wall and associated $P'$ anomaly 
necessarily and sufficiently classify the global anomalies.
For lower dimensions than 6, we find only mod 2 type indices as is in the $SU(2)$ anomaly and
our $\mu$ domain wall looks appropriately detecting them.
In higher dimensions, however, we have more non-trivial indices, for example,  
$\pi_6(SU(2))=\mathbb{Z}_{12}$.
We do not understand how it appears when we extend our formulation to 8 dimensions or higher.
More mathematically precise treatment of our system would be required
to fully understand this.

Another interesting possibility is to use a simple pseudoscalar operator,
which was studied in a previous work by Neuberger \cite{Neuberger:2003yg}
\begin{eqnarray}
i\mu \epsilon(x_5) \gamma_7,
\end{eqnarray}
which is a {\it twisted} mass under $U(1)_A$ rotation.
The fermion determinant 
\begin{eqnarray}
\det \left(\frac{D^{\rm 6D}+M\epsilon(x_6)
+i\mu\epsilon(x_5)\gamma_7  }{D^{\rm 6D}+M+i\mu\gamma_7  }\right),
\label{eq:detDWgamma7}
\end{eqnarray}
has a single Weyl fermion mode in the low-energy limit, too.
However,  as the pseudoscalar operator is odd in either of 
$P'$ and ($\pi$ rotation of ) $U(1)_A$,
both of the two domain walls produce the $CS$ terms and
the relation to the global anomaly is unclear.

The detailed mechanism of possible unphysical cancellations of the complex phase of the fermion determinant,
and how to choose the appropriate domain wall operators need a further investigation.

\section{A proposal for lattice regularization}
\label{sec:lattice}

Since our formulation is based on a massive Dirac fermion 
in 6-dimensions, it is natural to assume that a non-perturbative 
lattice regularization using the Wilson fermion is avaiblable, 
as it shares the same symmetries as in the continuum formulation.
Here we just give a simple proposal how to regularize 
our 6-dimensional Dirac fermion system on a lattice.
Detailed analysis about locality of the resulting 4-dimensional
theory, decoupling the doublers, {\it modified} chiral gauge symmetry, 
{\it etc}. will be discussed elsewhere.

First we pick up a set of link variables $\{U_\mu(\bar{x})\}(\mu=1,\cdots 4)$
on the 4-dimensional junction at $(x_5,x_6)=(0,0)$.
Then we solve the lattice version of the Yang-Mills gradient flow equation,
\begin{eqnarray}
\frac{\partial}{\partial t}U^t_\mu(\bar{x})=-\left\{\partial_{x,\mu}S_G(U^t)\right\}U^t_\mu(\bar{x}),
\end{eqnarray}
using $U^0_\mu(\bar{x})=U_\mu(\bar{x})$ as the initial condition, 
where $\partial_{x,\mu}S_G(U^t)$ denotes the Lie derivative of
the gauge action $S_G(U^t)$ with respect to $U^t_\mu(\bar{x})$, to define
\begin{equation}
U_\mu(\bar{x},x_5,x_6)=U^{|x_5|+|x_6|}_\mu(\bar{x}).
\end{equation}
Here we always set $U_5=U_6=$unity.
Note that the resulting link variables $U_\mu(\bar{x},x_5,x_6)$ are 
symmetric under $x_5\to -x_5$ and $x_6\to -x_6$.


We are now ready to {\it define} the 4-dimensional path integral of
anomaly free theory with Weyl fermions.
Together with the gauge part of the action $S_G(\{U_\mu(\bar{x})\})$,
we define
\begin{eqnarray}
\int DU_\mu(\bar{x}) e^{-S_G(\{U_\mu(\bar{x})\})}\prod_{i} \exp \left[-W_{\rm lat}^i(\{U_\mu(\bar{x})\})\right],
\end{eqnarray}
where 
\begin{eqnarray}
\exp \left[-W_{\rm lat}^i(\{U_\mu(\bar{x})\})\right]=\hspace{2in}\nonumber\\
\det \left(\frac{D_W^{{\rm 6D}R_i}+M_i\epsilon(x_6-a/2)\epsilon(L_6-x_6-a/2)
+i\mu_i\epsilon(x_5-a/2)\epsilon(L_5-x_5-a/2)
\gamma_6\gamma_7 }{D_W^{{\rm 6D}R_i}+M_i+i\mu_i\gamma_6\gamma_7 }\right),
\nonumber\\
\end{eqnarray}
where $D_W^{{\rm 6D}R_i}$ denotes the Wilson Dirac operator
in the $R_i$ representation of the gauge group, and 
$M_i$ and $\mu_i$ are chosen to be positive/negative for positive/negative chiral modes.
Note that the Wilson term has to have an opposite sign to $M_i$ and $\mu_i$.
These mass parameters are to be of the order of the lattice cut-off $1/a$.
However, to avoid contamination from the doubler modes, $M_i$ and $\mu_i$  
should have upper bounds, too.

In the above formula, the argument of the sign functions is shifted by $-a/2$ with 
the lattice spacing $a$ so that it is well-defined on 
integer values of coordinates on the lattice.
We always assume that the set of fermion flavors satisfy the anomaly free conditions
Eqs.~(\ref{eq:pfree}) and (\ref{eq:gfree}).

As a final remark of this section, 
we note that the full chiral gauge symmetry will  not be satisfied until 
we take the $L_5=L_6=\infty$ limits.


\section{Summary and discussion}
\label{sec:summary}

We have proposed a 6-dimensional regularization of
the chiral gauge theories in 4-dimensions.
Using the two different kinds of domain-walls, 
we have succeeded in localizing a single Weyl fermion at the
junction of the domain-walls.
One domain-wall is made giving
a kink mass in the 6-th direction to the fermions,
while another domain-wall is made by giving a kink structure in the 5-th direction 
to a background operator which is insensitive to the $U(1)_A$ rotation.

The conventional $M$ domain-wall mediates the perturbative anomaly inflow
and naturally exhibits the chain of the 6-dimensional $U(1)_A$, 
5-dimensional parity, and 4-dimensional gauge anomalies,
known as the descent equations
found by Stora \cite{Stora:1983ct} 
and Zumino \cite{Zumino:1983ew, Zumino:1983rz}.
On another domain-wall, the fermions are forced to form
(almost) a real representation and only mediates the mod-two type anomaly,
which we have assumed to be the source of the global anomalies.

The anomaly free condition of the target 4-dimensional gauge theory
is translated to the one for the set of 6-dimensional Dirac fermion
determinants to keep the axial $U(1)$ and $P'$ symmetries.
Using the Yang-Mills gradient flow in the 5-th and 6-th directions,
we can control the remnant of the gauge non-invariance due to the finite cut-offs,
and decouple the Weyl fermions at the junctions of anti-domain-walls.
As our formulation is nothing but a massive vector-like theory,
we expect that a non-perturbative regularization on a lattice is possible,
using standard Wilson Dirac fermions.

There are still a lot of open issues to be investigated.
There is an arbitrariness in the choice of the $\mu$ domain-wall operator,
to realize a single Weyl fermion at the domain-wall junction.
It is also unclear if the $\mu$ domain-wall and associated $P'$ anomaly 
necessarily and sufficiently classify the global anomalies.

In even dimensions, the $P'$ symmetry and its anomaly are usually neglected.
Our work, however, suggests its relation to the global anomalies in lower dimensions.
If we can formulate the $P'$ anomaly on a lattice, the lattice Dirac operator
could have a {\it modified} $P'$ symmetry, analogous to
the {\it modified} chiral symmetry \cite{Luscher:1998pqa} through 
the Ginsparg-Wilson relation \cite{Ginsparg:1981bj}.
It is an interesting question if the modified Dirac operator
realizes the exotic mod-$n$ index theorems, 
identifying explicit link variable configurations 
which give non-trivial indices on the lattice.

In our formulation, we have switched off the gauge fields 
in the directions of extra-dimensions and use the Yang-Mills gradient flow
to maintain the 4-dimensional gauge invariance.
One concern is that this treatment of the gauge fields is non-local
in the extra-dimensions and may not fully decouple the mirror fermions,
which was already discussed in Ref.~\cite{Grabowska:2015qpk}.
It is then an interesting question if our formulation can be
extended to a model with physical 
extra dimensions also in the gauge sector.
Such a direction may be linked to studies of higher dimensional 
beyond the standard models \cite{Asaka:2001eh}.

Our formulation suggests that there is a possibility
of doubly gapped topological insulators in four-dimensions, 
having a conducting mode on two-dimensional edges,
which may be realized in condensed matter systems.

Finally, it would be great if we can incorporate the Higgs field 
to our 6-dimensional lattice and give
{\it a non-perturbative definition} of the standard model,
which is also an interesting subject for further study.


\section*{Acknowledgments}
We thank S.~Aoki, D.~Grabowska, D.~B.~Kaplan, Y.~Kikukawa, H.~Suzuki, Y.~Tachikawa, 
and S.~Yamaguchi for useful discussions.
We also thank K.~Hashimoto for organizing a study group on topological insulators,
which helped our work a lot.
This work is supported in part by the Grand-in-Aid of the Japanese Ministry of Education 
No.25800147, 26247043 (H.F.), No. 26400248(T.O.), and No. 15J01081 (R.Y.).

\appendix

\section{Gamma matrices}
\label{app:gamma}

Although our results do not depend on the 
basis of the gamma matrices,
we summarize here the most convenient one
to make our analysis simple.

For the Euclidean 4-dimensional gamma matrices, we use
the so-called chiral representation:
\begin{eqnarray}
\bar{\gamma}_{i=1,2,3} &=&
\left(
\begin{array}{cc}
 & -i\sigma_i\\
i\sigma_i&
\end{array}
\right),
\;\;\;
\bar{\gamma}_4=\left(
\begin{array}{cc}
 & \mathbb{I}\\
\mathbb{I}&
\end{array}
\right),\;\;\;
\end{eqnarray}
where $\sigma_i$ denote the Pauli matrices, and
$\mathbb{I}$ is the $2\times 2$ identity matrix.

In this paper, we also introduce another set of the gamma matrices,
\begin{eqnarray}
\bar{\gamma}'_{i=1,2,3,4} = i\bar{\gamma}_5\bar{\gamma}_i, \;\;\;\bar{\gamma}_5'=\bar{\gamma}_5.
\end{eqnarray}
Note that the matrices $\bar{\gamma}'_i$ satisfy the same Clifford algebra as $\bar{\gamma}_i$.

For the $8\times 8$ gamma matrices in 6-dimensions, we use 
\begin{eqnarray}
\gamma_{i=1,2,3,4} &=&
\left(
\begin{array}{cc}
\bar{\gamma}_i & \\
&-\bar{\gamma}_i
\end{array}
\right),
\;\;\;
\gamma_5=\left(
\begin{array}{cc}
 & -i\mathbb{I}\\
i\mathbb{I}&
\end{array}
\right),\;\;\;
\gamma_6=\left(
\begin{array}{cc}
 & \mathbb{I}\\
\mathbb{I}&
\end{array}
\right),\;\;\;
\end{eqnarray}
where $\mathbb{I}$ is the $4\times 4$ identity matrix.

With these gamma matrices, the chiral operators are given as
\begin{eqnarray}
\bar{\gamma}_5 = 
\left(
\begin{array}{cc}
\mathbb{I}&\\
&-\mathbb{I}
\end{array}
\right),\;\;\;
\gamma_7 = 
\left(
\begin{array}{cc}
\bar{\gamma}_5&\\
&-\bar{\gamma}_5
\end{array}
\right).
\end{eqnarray}
It is also useful to note that $i\gamma_5\gamma_6\gamma_7$ is represented by
\begin{eqnarray}
i\gamma_5\gamma_6\gamma_7 =
\left(
\begin{array}{cc}
\bar{\gamma}_5&\\
&\bar{\gamma}_5
\end{array}
\right),
\end{eqnarray}
so that one can easily confirm that the constraints $\gamma_6=\pm 1$
and $i\gamma_5\gamma_6\gamma_7 =\pm 1$ on the 6-dimensional spinor,
lead to $\bar{\gamma_5}=\pm 1$ on the 4-dimensional spinor.

\section{Bulk/edge decomposition of the 5-dimensional domain-wall fermion determinant}
\label{app:bulkedge}

It was shown a long ago by Callan and Harvey \cite{Callan:1984sa} that
the 5-dimensional domain-wall fermion determinant can be 
decomposed into the bulk part, which produces the CS term,
and the edge part, which converges to the Weyl fermion determinant,
canceling the gauge non-invariance with each other.
However, there has been no explicit formula for the decomposition,
except for the one at one-loop level \cite{Chandrasekharan:1993ag}.
Here we propose a non-perturbative method for the bulk/edge decomposition.

The difficulty in the decomposition is in the fact that
we have to introduce the gauge non-symmetric regulator
to separate the bulk and edge modes.
For example, if we introduce a simple mass $\mu_2$ for this,
\begin{eqnarray}
\det\left(\frac{\bar{D}^{\rm 5D}+\mu\epsilon(x_5)\epsilon(L_5-x_5)}{\bar{D}^{\rm 5D}+\mu}\right)&=&
\det\left(\frac{\bar{D}^{\rm 5D}+\mu\epsilon(x_5)\epsilon(L_5-x_5)+\mu_2}{\bar{D}^{\rm 5D}+\mu}\right)
\nonumber\\&&
\times\det\left(\frac{\bar{D}^{\rm 5D}+\mu\epsilon(x_5)\epsilon(L_5-x_5)}{\bar{D}^{\rm 5D}+\mu\epsilon(x_5)\epsilon(L_5-x_5)+\mu_2}\right),
\end{eqnarray}
we end up with a Weyl fermion determinant, which produces the so-called covariant anomaly.
This means that the decomposition is not complete but the high energy modes 
still have a part of the boundary effective action
which compensates the difference between the consistent and covariant anomaly.

Here we introduce a {\it mass term} which breaks the gauge symmetry
only at the boundaries $x_5=0$ and $x_5=L_5$:
\begin{eqnarray}
\mu_2[\bar{\psi}(\bar{x},0)\psi(\bar{x},L_5)+\bar{\psi}(\bar{x},L_5)\psi(\bar{x},0)],
\end{eqnarray}
where $\bar{x}=(x_1,x_2,x_3,x_4)$.
Note that this is the conventional mass term used in the domain-wall fermions 
in the vector-like theories.
The fermion action with this mass term is rewritten as
\begin{eqnarray}
S_F = \int d^5 x \int d^5x' \bar{\psi}(\bar{x},x_5)\left[\delta(x-x')\left\{\bar{D}^{\rm 5D}+\mu\epsilon(x_5)\epsilon(L_5-x_5)\right\}+\mu_2^{x_5,x_5'} \right]\psi(x'),
\end{eqnarray}
where 
\begin{equation}
\mu_2^{x_5,x_5'}\equiv \mu_2 \left[
\delta(x_5)\delta(x_5'-L_5)+\delta(x_5-L_5)\delta(x_5')
\right],
\end{equation}
and our target fermion determinant with the Pauli-Villars fields can be decomposed as
\begin{eqnarray}
\det\left(\frac{\bar{D}^{\rm 5D}+\mu\epsilon(x_5)\epsilon(L_5-x_5)}{\bar{D}^{\rm 5D}+\mu}\right)&=&
{\rm Det}\left(\frac{\delta(x-x')(\bar{D}^{\rm 5D}+\mu\epsilon(x_5)\epsilon(L_5-x_5))+\mu_2^{x_5,x_5'}}{\delta(x-x')(\bar{D}^{\rm 5D}+\mu)}\right)
\nonumber\\&&
\times{\rm Det}\left(\frac{\delta(x-x')(\bar{D}^{\rm 5D}+\mu\epsilon(x_5)\epsilon(L_5-x_5))}{\delta(x-x')(\bar{D}^{\rm 5D}+\mu\epsilon(x_5)\epsilon(L_5-x_5))+\mu_2^{x_5,x_5'}}\right),
\nonumber\\
\label{eq:decomposition}
\end{eqnarray}
where the determinant ${\rm Det}$ is taken in the doubled space of $x$ and $x'$.

To the second determinant, only boundary Weyl fermion modes with positive chirality at $x_5=0$
and negative chirality at $x_5=L_5$ contribute so that
\begin{eqnarray}
\lim_{\mu\to\infty}
{\rm Det}\left(\frac{\delta(x-x')(\bar{D}^{\rm 5D}+\mu\epsilon(x_5)\epsilon(L_5-x_5))}{\delta(x-x')(\bar{D}^{\rm 5D}+\mu\epsilon(x_5)\epsilon(L_5-x_5))+\mu_2^{x_5,x_5'}}\right)
= \det\frac{\mathcal{D}}{\mathcal{D}+\mu_2},
\end{eqnarray}
holds, where $\mathcal{D}$ is defined as
\begin{eqnarray}
\mathcal{D} = P^5_-\bar{D}^{\rm 4D}P^5_+ + P^5_+\bar{\partial}^{\rm 4D}P^5_-,
\end{eqnarray}
with $\bar{D}^{\rm 4D}=\sum_{i=1}^4\bar{\gamma}'_i\nabla_i|_{x_6=x_5=0}$, 
$\bar{\partial}^{\rm 4D}=\sum_{i=1}^4\bar{\gamma}'_i\nabla_i|_{x_6=0,x_5=L_5}$ and $P^5_\pm=(1\pm \bar{\gamma_5})/2$.
This form of the fermion determinant with Pauli-Villars is known to correctly produce
the consistent anomaly. 
This justifies a naive computation of the imaginary part of the
first determinant in Eq.~(\ref{eq:decomposition}), which leads to $\pi CS^{(x_5<0)}$.


\section{Fermion determinant on the $\mu$ domain-wall}
\label{app:muDWdetails}

In this appendix, we give the details of
the computation in Eqs.~(\ref{eq:5Ddetmu}) and (\ref{eq:4thdetmuDW}).

For this purpose, it is enough to consider 
\begin{equation}
\label{eq:detorg}
\lim_{\mu \to \infty}\det \left(\frac{D^{\rm 6D}+i\mu\epsilon(x_5)\gamma_6\gamma_7 +M_1}
{D^{\rm 6D}+i\mu\epsilon(x_5)\gamma_6\gamma_7 +M_2}\right),
\end{equation}
in the $\mu\to \infty$ limit with arbitrary masses $M_1$ and $M_2$.
It receives contributions only from the boundary localized modes,
which are constrained to satisfy
\begin{eqnarray}
\gamma_5(\partial_5+i\mu\epsilon(x_5)\gamma_5\gamma_6\gamma_7  )\psi =0,
\end{eqnarray}
whose solution is given by
\begin{eqnarray}
\psi = e^{-\mu |x_5|}\psi',\;\;\; i\gamma_5\gamma_6\gamma_7 \psi'=\psi'.
\end{eqnarray}

The operator $i\gamma_5\gamma_6\gamma_7 $ has a $4\times 4$ 
block-diagonal form so that its projection operator can be expressed as
\begin{eqnarray}
\hat{P}^6_\pm \equiv(1\pm i\gamma_5\gamma_6\gamma_7 )/2
=\left(
\begin{array}{cc}
P_\pm\\
& P_\pm
\end{array}
\right),
\end{eqnarray}
where $P_\pm \equiv (1\pm\bar{\gamma}_5)/2$ are 
projection operators for 4-component spinors.

With the above constraint, multiplying $-i\gamma_5$,
and denoting
$D^{\rm 5D}=(\sum_{i=1}^4\gamma_i 
\nabla_i+\gamma_6\partial_6)|_{x_5=0}$, the determinant Eq.~(\ref{eq:detorg})
can be rewritten as
\begin{eqnarray}
\det \left[\hat{P}^6_+\left(-i\gamma_5
D^{\rm 5D}-i\gamma_5 M_2\right)^{-1}
\left(-i\gamma_5 D^{\rm 5D}-i\gamma_5 M_1\right)\hat{P}^6_++\hat{P}^6_-\right].
\end{eqnarray}
Inserting the two unitary operators
\begin{eqnarray}
Q_1 = \frac{1}{\sqrt{2}}
\left(
\begin{array}{cc}
i\bar{\gamma}_5&-i\mathbb{I}\\
\mathbb{I} & \bar{\gamma}_5
\end{array}
\right),\;\;\;
Q_2 = \frac{1}{\sqrt{2}}
\left(
\begin{array}{cc}
-i\mathbb{I} & -i\bar{\gamma}_5\\
-\bar{\gamma}_5&\mathbb{I}
\end{array}
\right),\;\;\;
\end{eqnarray}
we obtain a $4\times 4$ block-diagonalized form
\begin{eqnarray}
\det \left[\hat{P}^6_+Q_1^\dagger Q_1\left(-i\gamma_5
D^{\rm 5D}-i\gamma_5 M_2\right)^{-1}
Q_2^\dagger Q_2\left(-i\gamma_5 D^{\rm 5D}-i\gamma_5 M_1\right)Q_1^\dagger Q_1\hat{P}^6_++\hat{P}^6_-\right]
\nonumber\\
= \det \left[\hat{P}^6_+Q_1^\dagger \left(
\begin{array}{cc}
(\hat{D}^{\rm 5D}-M_2)^{-1}(\hat{D}^{\rm 5D}-M_1) & \\
& (\hat{D}^{\rm 5D}+M_2)^{-1}(\hat{D}^{\rm 5D}+M_1)
\end{array}
\right)Q_1\hat{P}^6_++\hat{P}^6_-\right],
\nonumber\\
\end{eqnarray}
where $\hat{D}^{\rm 5D}=(\sum_{i=1}^4\bar{\gamma}_i \nabla_i+\bar{\gamma_5}\partial_6)|_{x_5=0}$.
Since $Q_1$ commutes with $\hat{P}^6_+$, we can factorize the
determinant as
\begin{eqnarray}
\label{eq:det2decomposition}
\det \left[P_+ (\hat{D}^{\rm 5D}+M_2)^{-1}(\hat{D}^{\rm 5D}+M_1) P_++P_-\right]
\nonumber\\
\times \det \left[P_+ (\hat{D}^{\rm 5D}-M_2)^{-1}(\hat{D}^{\rm 5D}-M_1) P_++P_-\right].
\end{eqnarray}
When $M_1$ and $M_2$ both commute with $\hat{D}^{\rm 5D}$,
this determinant is not only real but positive.

\bibliographystyle{ptephy}
\bibliography{bib}

\end{document}